\documentclass[reprint, amsmath, amssymb, aps]{revtex4-1}
\usepackage{graphicx}
\usepackage{dcolumn}
\usepackage{bm}
\usepackage{mathrsfs}
\usepackage{hyperref}
\usepackage{epstopdf}
\usepackage{amsmath}
\usepackage{amssymb}
\usepackage{mathrsfs}
\usepackage{diagbox}
\usepackage{booktabs}
\usepackage{multirow}
\usepackage{makecell}

\begin{document}
\title{Long-wavelength emergent phonons in skyrmion crystals distorted by exchange anisotropy and tilted magnetic fields}
\author{Yangfan Hu}
 \email[Corresponding author.]{huyf3@mail.sysu.edu.cn}
\affiliation{Sino-French Institute of Nuclear Engineering and Technology, Sun Yat-sen University, 519082, Zhuhai, China}

\begin{abstract}
    Skyrmion crystals (SkX) are periodic alignment of magnetic skyrmions, i.e., a type of topologically protected spin textures. Compared with ordinary crystals, they can be drastically deformed under anisotropic effects because they are composed of field patterns whose deformation does not cause any bond-breaking. This exotic ductility of SkX bring about great tunability of its collective excitations called emergent phonons, which are vital for magnonics application. The question is how to quantitatively determine the emergent phonons of distorted SkX. Here we systematically study the long wavelength emergent phonons of SkX distorted by (a) a negative exchange anisotropy, and (b) a tilted magnetic field. In both cases, deformation and structural transitions of SkX thoroughly influence the frequency, anisotropy of vibrational pattern and dispersion relation, and coupling between lattice vibration and in-lattice vibration for all modes. Tilted magnetic fields are very effective in tuning the emergent phonons, such that all modes except the Goldstone mode can be excited by AC magnetic fields when a tilted bias field is presented.
\end{abstract}

\maketitle

SkX\cite{1,2,3} are a type of emergent crystalline states made up of magnetic skyrmions\cite{4,5}, i.e., a type of spin solitons with nontrivial topology, appearing in magnetic materials. They are important for realizing bottom-up microwave- and magnonics-related application\cite{6} that go beyond state-of-the-art nanotechnology, because of their spontaneous existence\cite{5,7}, nanometer-sized composing “particles”\cite{7,8} , and exotic robustness\cite{3,9} due to local topological protection of skyrmions and global protection induced by mode-mode interactions\cite{10}.. Moreover, an intrinsic advantage of SkX over any artificial superlattices is that unlike the latter ones which have almost fixed atomic structures once designed, SkX can undergo drastic deformation\cite{11,12,45} or even structural phase transitions\cite{3,13} when subject to various kinds of external fields\cite{11,12,14,15}. Since the collective excitations of crystals depend sensitively on deformation and transition of the structure, we are expected to achieve great tunability of the dynamic properties of SkX while keeping the underlying material unchanged. The key scientific problem behind is how to quantitatively determine the elementary excitations of a distorted SkX phase under different conditions external fields. The collective excitations of SkX has been extensively studied\cite{6,16,17,18,19,20,21,22,23,24}, yet the influence of anisotropy on them has never been clarified.

Practically, effects of anisotropy is not only important, but also inevitable, because presence of various kinds of anisotropic interactions are naturally permitted by the symmetry of helimagnets\cite{25,26,27}. These intrinsic anisotropies are closely related to the intrinsic distortion of skyrmion lattice revealed by the SANS experiment in MnSi\cite{1} and and $Cu_2OSeO_3$\cite{15}, as well as the triangular-square structural transition of SkX observed in $Co_8Zn_8Mn_4$\cite{3} and MnSi\cite{13}. The significance of this structural anisotropy to the elementary excitations has already been observed when studying the spin excitations of SkX in $Cu_2OSeO_3$\cite{28}. Meanwhile, in a previous work\cite{29} we have find that the emergent phonons of undeformed hexagonal Bloch-type SkX at long wavelength limit are fundamentally different from those of ordinary crystals, such that the lattice vibration (resembling the acoustical branches of ordinary phonons) and in-lattice vibration (resembling the optical branches of ordinary phonons) of SkX are coupled in a proportion of all modes at long wavelength limit, leading to appearance of various types of “emergent elastic waves” with finite frequencies. It is of fundamental interest to see how such a unique feature of SkX changes when the distortion of SkX comes into play.

In this work we study the emergent phonons in distorted Bloch-type SkX near the long wavelength limit under two types of conditions: a) the distortion is induced by the intrinsic exchange anisotropy of the material, and b) the distortion is induced by a tilted bias magnetic field with an in-plane component. In both cases studied, we find that the deformation and structural transition of SkX has a thorough influence on the frequency, mode of vibrational, anisotropy of dispersion relation, and coupling between lattice vibration and in-lattice vibration for all emergent phonons. Specifically, lattice vibration and in-lattice vibration are always coupled at long wavelength limit for all the emergent phonon modes in SkX distorted by a tilted magnetic field. Moreover, when no anisotropic effects are consider, about half of the emergent phonon modes cannot be excited by applying a AC magnetic field. We find that when a tilted magnetic field is applied, all modes except the Goldstone mode can all be excited by an AC magnetic field.

\section{Results and Discussion}
\
\\
\\
$Model$ \\
For deformable Bloch-type SkX in B20 chiral magnets, we use the following Landau-Ginzburg functional to describe the rescaled free energy density of the system (rescaling process shown in the Methods section) \cite{1}
\begin{equation}
\begin{aligned}
    \tilde{\phi }\left( \mathbf{m} \right)=&\sum\limits_{i=1}^{3}{{{\left( \frac{\partial \mathbf{m}}{\partial {{r}_{i}}} \right)}^{2}}+2}\mathbf{m}\cdot \left( \nabla \times \mathbf{m} \right)-2\mathbf{b}\cdot \mathbf{m}\\&+\tilde{T}{{\mathbf{m}}^{2}}+{{\mathbf{m}}^{4}}+{{\tilde{A}}_{e}}\sum\limits_{i=1}^{3}{{{\left( \frac{\partial {{m}_{i}}}{\partial {{r}_{i}}} \right)}^{2}}},
\label{1}
\end{aligned}
\end{equation}
where $\mathbf{m}$ is the rescaled magnetization, $\mathbf{b}$ is the rescaled magnetic field, $\tilde{T}$ is the rescaled temperature, and ${{\tilde{A}}_{e}}$ is the rescaled exchange anisotropy. In this work we study the following two types of conditions: (a) $\mathbf{b}={{\left[ \begin{matrix}
    0 & 0 & b  \\
 \end{matrix} \right]}^{T}}$ and ${{\tilde{A}}_{e}}<0$; (b) $\mathbf{b}={{\left[ \begin{matrix}
    \frac{\sqrt{2}}{2}b\sin \theta  & \frac{\sqrt{2}}{2}b\sin \theta  & b\cos \theta   \\
 \end{matrix} \right]}^{T}}$ and ${{\tilde{A}}_{e}}=0$, where $\theta $ denotes the angle between the  $z$-axis and the direction of the applied magnetic field, and when $\theta =\frac{\pi }{2}$ the external magnetic field is applied along [110]. The two conditions (a) and (b) correspond to the simplest cases where the effect of an intrinsic anisotropic interaction and the effect of an anisotropic external field are considered, respectively. Condition (a) is chosen because it is previously understood\cite{27} that a negative exchange anisotropy not only explains the intrinsic anisotropy of SkX reflected by the unequal intensities of the six Bragg spots in the SANS experiment of $MnSi$\cite{1} and $Cu_2OSeO_3$\cite{15}, but also repeat the triangle-square structural transition of SkX observed in $Co_8Zn_8Mn_4$\cite{3} and MnSi\cite{13}. Condition (b) is chosen because a tilted bias magnetic field is probably the most convenient external field\cite{14,30,31,32,33} to induce a distortion of SkX, and a bias magnetic field is sometimes required to obtain a thermodynamically stable SkX\cite{1}. Moreover, it may induce an easy axis (or hard axis) for the current driven motion of skyrmions\cite{34,35,36,37} in the SkX phase, whose direction is tunable by the in-plane direction of the tilted magnetic field. A tunable anisotropy of the skyrmion Hall effect\cite{38,39,40,41} is also anticipated. 

Deformable SkX with long range order can be expressed analytically by the following Fourier expansion of\cite{10,42} 
\begin{equation}
    \begin{aligned}
        \mathbf{m}=\sum\limits_{\mathbf{l}}{{{\mathbf{m}}_{{{\mathbf{q}}_{\mathbf{l}}}}}{{e}^{\text{i}{{\mathbf{q}}^e_{\mathbf{l}}}(\varepsilon _{ij}^{e},\ {{\omega }^{e}})\cdot \mathbf{r}}}},
        \label{2}
    \end{aligned}
\end{equation}
where ${{\mathbf{m}}_{{{\mathbf{q}}_{\mathbf{l}}}}}$ denotes the Fourier magnitudes, ${{\mathbf{q}}^e_{\mathbf{l}}}={{l}_{1}}{{\mathbf{q}}^e_{\mathbf{1}}}+{{l}_{2}}{{\mathbf{q}}^e_{2}}$ where ${{l}_{1}}$ and ${{l}_{2}}$ are integers and ${{\mathbf{q}}^e_{\mathbf{1}}}$ and ${{\mathbf{q}}^e_{2}}$ are the basic reciprocal vectors of SkX, which are deformable under external disturbance. The deformation of ${{\mathbf{q}}^e_{\mathbf{1}}}$ and ${{\mathbf{q}}^e_{2}}$ is described by the emergent elastic strains $\varepsilon _{ij}^{e}$ and the emergent rotational angle ${{\omega }^{e}}$\cite{42}, which are defined from the emergent displacement field ${{\mathbf{u}}^{e}}={{\left[ \begin{matrix}
       u_{1}^{e} & u_{2}^{e}  \\
\end{matrix} \right]}^{T}}$ by $\varepsilon _{ij}^{e}=\frac{1}{2}(u_{i,j}^{e}+u_{j,i}^{e})$ and ${{\omega }^{e}}=\frac{1}{2}(u_{1,2}^{e}-u_{2,1}^{e})$. Similar to atomic lattice, rigid translation of SkX does not induce a change of free energy, thus it is $\varepsilon _{ij}^{e}$ and ${{\omega }^{e}}$ instead of ${{\mathbf{u}}^{e}}$ which appear in the expression of the equilibrium magnetization. The expression of ${{\mathbf{q}}^e_{\mathbf{l}}}$ in terms of $\varepsilon _{ij}^{e}$ and ${{\omega }^{e}}$ depends on the crystalline structure of SkX, and is introduced in the Methods section for hexagonal SkX.
    
For condition (a), minimization of the free energy based on eqs. (1, 2) at given $b$, $\tilde{T}$ and ${{\tilde{A}}_{e}}$ determines if there is a metastable SkX phase. For a metastable SkX phase to become thermodynamically stable, the minimized free energy for the SkX phase must be the smallest one among all considered phases such as the ferromagnetic phase and the generalized conical (G-conical) phase (the G-conical phase is described by a constant magnetization vector plus a helix of magnetization, where the direction of both vectors is free to rotate in space). For condition (b), similar calculation is performed at given $b$, $\tilde{T}$ and $\theta $. One should notice that in condition (b) we assume the 2D SkX is always distributed in the x-y plane with the tilting of magnetic field. In reality this corresponds to SkX in a magnetic thin film with thickness less than a period of the SkX\cite{43}. For the same reason, the direction of magnetization helix of the G-conical phase is fixed in the SkX plane so that we have the in-plane single-Q (IPSQ)\cite{10} phase instead of the G-conical phase in Figure 1(g). The ${{\tilde{A}}_{e}}-b$ phase diagram for condition (a) and the $\theta -b$ phase diagram for condition (b) calculated at $\tilde{T}=0.5$ are plotted in FIG. 1(f) and 1(g), respectively. In the two phase diagrams, the stable SkX region is marked blue while the metastable SkX region corresponds to a shadowed light green area. Due to the exotic robustness of the SkX phase confirmed in experiments\cite{3,9}, we focus on the metastable SkX here. The decrease of ${{\tilde{A}}_{e}}$ leads to a compression of the lattices of SkX in the $y$ direction, and eventually leads to a triangle-square structural phase transition\cite{27} (FIG. 1(a-c)). On the other hand, the increase of $\theta $ leads to a rotation of SkX in the $x-y$ plane, such that a pair of the vertexes of hexagon aligns with the in-plane direction of the magnetic field(FIG. 1(a, d, e)). Meanwhile, existence of an in-plane magnetic field breaks the hexagonal symmetry of the field configuration of skyrmion: as illustrated in FIG. 1(e), the magnitude of in-plane components of magnetization inside the pink circle region is depressed since they align opposite to the in-plane magnetic field, while that inside the green circle region is enhanced.

For a metastable SkX state of interest, its equilibrium magnetization is described by ${{\left( {{\mathbf{m}}_{{{\mathbf{q}}_{\mathbf{l}}}}} \right)}_{st}}$, ${{\left( \varepsilon _{ij}^{e} \right)}_{st}}$ and ${{\left( {{\omega }^{e}} \right)}_{st}}$, where the latter two determine a equilibrium emergent displacement field ${{\left( {{\mathbf{u}}^{e}} \right)}_{st}}$. Consider a small vibration around this metastable state, which induces simultaneously a vibration of ${{\mathbf{u}}^{e}}$ denoted by ${{\left( {{\mathbf{u}}^{e}} \right)}_{v}}(\mathbf{r},t)$ and a vibration of ${{\mathbf{m}}_{{{\mathbf{q}}_{\mathbf{l}}}}}$ denoted by ${{\left( {{\mathbf{m}}_{{{\mathbf{q}}_{\mathbf{l}}}}} \right)}_{v}}(\mathbf{r},t)$. In this case, eq. (2) becomes

\begin{equation}
\begin{aligned}
\mathbf{m}=\sum\limits_{\mathbf{l}}{\left[ {{\left( {{\mathbf{m}}_{{{\mathbf{q}}_{\mathbf{l}}}}} \right)}_{st}}+{{\left( {{\mathbf{m}}_{{{\mathbf{q}}_{\mathbf{l}}}}} \right)}_{v}} \right]{{e}^{\text{i}\left({\mathbf{q}}^e_{\mathbf{l}}\right)_{st}\cdot \left[ \mathbf{r}-{{\left( {{\mathbf{u}}^{e}} \right)}_{v}} \right]}}},
\label{3}
\end{aligned}
\end{equation}
where $\left({\mathbf{q}}^e_{\mathbf{l}}\right)_{st}={\mathbf{q}}^e_{\mathbf{l}}\left[ {{\left( \varepsilon _{ij}^{e} \right)}_{st}},\ {{\left( {{\omega }^{e}} \right)}_{st}} \right]$. It is convenient to write components of all Fourier magnitudes ${{\mathbf{m}}_{{{\mathbf{q}}_{\mathbf{l}}}}}$ in a single vector ${{\mathbf{m}}^{q}}$, for which the two vectors ${{\left( {{\mathbf{u}}^{e}} \right)}_{v}}$ and ${{\left( {{\mathbf{m}}^{q}} \right)}_{v}}$  include all the variables to be solved. 

The dispersion relation for the coupled wave motion of ${{\left( {{\mathbf{u}}^{e}} \right)}_{v}}$ and ${{\left( {{\mathbf{m}}^{q}} \right)}_{v}}$ can be obtained by considering the plane-wave form of solution ${{\left( {{\mathbf{u}}^{e}} \right)}_{vib}}={{\mathbf{u}}^{e0}}{{e}^{\text{i}(\mathbf{\tilde{k}}\cdot \mathbf{r}-\omega t)}}$, ${{\left( {{\mathbf{m}}^{q}} \right)}_{vib}}={{\mathbf{m}}^{q0}}{{e}^{\text{i}(\mathbf{\tilde{k}}\cdot \mathbf{r}-\omega t)}}$ for the Euler-Lagrangian equation of them\cite{29,44} derived from the least action principle, which is briefly introduced in the Methods section. The details on how to derive the Euler-Lagrangian equation is introduced in \cite{44}, and the solution process of the dispersion relation is exactly the same as that introduced in \cite{29}. We denote the frequencies of different modes by ${{\omega }_{i}}(\mathbf{\tilde{k}})$ (or equivalently ${{\tilde{\omega }}_{i}}(\mathbf{\tilde{k}})={{\omega }_{i}}(\mathbf{\tilde{k}})/\eta $, where $\eta $ is a material dependent factor so that ${{\tilde{\omega }}_{i}}(\mathbf{\tilde{k}})$ is material independent), ordered in such a way that ${{\omega }_{1}}(\mathbf{0})<{{\omega }_{2}}(\mathbf{0})<{{\omega }_{3}}(\mathbf{0})<\cdots $ calculated at $\tilde{T}=0.5$, $b=0.3$, ${{\tilde{A}}_{e}}=0$ and $\theta =0$.
\
\\
\\
$Effects$ $of$ $exchange$ $anisotropy$ $on$ $the$ $emergent$ $phonons$ $of$ $SkX$ \\
We first consider the emergent phonons of metastable SkX at long wavelength limit under condition (a). In particular, we perform the calculation at $\tilde{T}=0.5$, $b=0.3$, and different values of ${{\tilde{A}}_{e}}$. As ${{\tilde{A}}_{e}}$ decreases, all modes present increasingly significant anisotropy of the vibrational pattern, which are generally categorized into two types of changes: (1) an induced anisotropy of the vibrational pattern of skyrmion inside the lattice, and (2) a change of vibrational pattern due to the triangle-square structural phase transition. Changes of type (1) includes: As ${{\tilde{A}}_{e}}$ decreases, the vibrational pattern of skyrmion in the ${{\omega }_{3}}$ (CCW) mode changes from a circle to a triangle (FIG. 2(a3, b3, $\cdots$, f3), Supplementary Videos a3, b3); one of the three “antenna” of the ${{\omega }_{4}}$ mode gradually disappears (FIG. 2(a4, b4, $\cdots$, f4) Supplementary Videos a4, b4); when extended to its maximum size during the breathing motion of ${{\omega }_{5}}$ mode, the skyrmion field pattern has two extremum points of ${{\mathbf{m}}_{3}}$ (FIG. 2(e5), Supplementary Videos a5, b5). Changes of type (2) includes: due to the triangle-square structural phase transition, the vibrational pattern of skyrmions of the ${{\omega }_{2}}$ mode will form parallel stripes, each of which similar to a Bloch-type domain wall, while new skyrmions are formed between every two stripes (FIG. 2(e2, f2), Supplementary Video b2); the shape of intermediate area between skyrmions for the ${{\omega }_{5}}$ (breathing) mode changes from a triangle (pink triangle in FiG. 2 (c5)) to a square (green square in FiG. 2 (e5)). 

Variation of the frequencies of the first 8 modes with ${{\tilde{A}}_{e}}$ is plotted in FiG. 3(a), where a break point appears for all modes at the triangle-square structural phase transition. Meanwhile, several neighboring modes (e.g., ${{\omega }_{4}}$ and ${{\omega }_{5}}$) undergo a crossing of frequency as ${{\tilde{A}}_{e}}$ decreases. 

The long wavelength dispersion relation of all modes possesses certain degree of anisotropy as ${{\tilde{A}}_{e}}$ decreases (FIG. 4), specifically, the dispersion relation ${{\omega }_{4}}(\mathbf{\tilde{k}})$ at long wavelength becomes nearly independent of ${{\tilde{k}}_{2}}$ (FIG. 4(c4)).
\
\\
\\
$Effects$ $of$ $a$ $tilted$ $magnetic$ $field$ $on$ $the$ $emergent$ $phonons$ $of$ $SkX$ \\
We then consider the emergent phonons of metastable SkX at long wavelength limit under condition (b). In particular, we perform the calculation at $\tilde{T}=0.5$, $b=0.3$, and different values of $\theta $ at long wavelength limit. As $\theta $ increases, the inter-mode coupling of the vibrational pattern becomes increasingly significant (FIG.2, Supplementary Videos c$i$, d$i$, ($i=1, 2, \cdots, 6$)). For instance, breathing vibration of ${{\mathbf{m}}_{3}}$ gradually appears for ${{\omega }_{2}}$, ${{\omega }_{3}}$, ${{\omega }_{4}}$ and ${{\omega }_{6}}$ as $\theta $ increases, indicating a coupling of these mode with ${{\omega }_{5}}$. Meanwhile, ${{\omega }_{5}}$ is strongly coupled with ${{\omega }_{6}}$ at large $\theta $ (e.g., $\theta ={{45}^{\circ }}$) such that the vibrational pattern of the two modes are similar to each other (FIG. 2(i5, j5, i6, j6), Supplementary Videos d5, d6). We also observe an increasing anisotropy of the vibrational pattern of all modes as $\theta $ increases, while the induced anisotropy depends on the direction of the in-plane magnetic field in a complicated way (FIG. 2, Supplementary Videos c$i$, d$i$, ($i=1, 2, \cdots, 6$)). 

The frequencies of all vibrational modes vary with $\theta $, as illustrated in FiG. 3(b). The long wavelength dispersion relation of all modes except ${{\omega }_{\text{1}}}$ (the Goldstone mode) possesses significant anisotropy as $\theta $ increases (FIG. 4), while the induced anisotropy depends sensitively on the direction of in-plane magnetic field. As illustrated in FIG. 4(e2, e3, e5), at $\theta ={{45}^{\circ }}$, the frequencies of ${{\omega }_{\text{2}}}$, ${{\omega }_{\text{3}}}$ and ${{\omega }_{\text{5}}}$ become almost independent of $\left| {\mathbf{\tilde{k}}} \right|$ near the   point when $\mathbf{\tilde{k}}$ is perpendicular to the in-plane magnetic field.
\
\\
\\
$Coupling$ $between$ $lattice$ $vibration$ $and$ $in-lattice$ $vibration$ $of$ $emergent$ $phonons$ $in$ $SkX$ $when$ $anisotropic$ $effects$ $are$ $considered$\\
In previous study\cite{29}, we have found that a unique feature of the emergent phonons at long wavelength limit is that a part of the modes permits coupling between vibration of the lattices and vibration of the field pattern inside the lattices. It means that multiple modes of emergent elastic waves with finite frequency at long wavelength limit are allowed to propagate in SkX. While anisotropic effects are considered, we find that generally speaking the coupling between vibration of the lattice and vibration of the skyrmion pattern inside the lattice is enhanced. To study the different effects induced by negative exchange anisotropy and tilted magnetic fields, we list in Table 1 the values of ${{{\mathbf{u}}^{e0}}}$ and ${{{\mathbf{m}}^{c0}}}$ in the eigenvectors for the first 16 modes of emergent phonons of SkX calculated at 5 different conditions of ${{\tilde{A}}_{e}}$ and $\theta $. As defined previously\cite{29}, the modes with nonzero components of ${{{\mathbf{u}}^{e0}}}$ in the eigenvector at long wavelength limit are called emergent elastic waves, and the modes with nonzero components of ${{{\mathbf{m}}^{c0}}}$ in the eigenvector can be stimulated by corresponding AC magnetic fields. As listed in Table1, presence of exchange anisotropy turns several modes to emergent elastic waves, while the values of $u_{1}^{e0}$ and $u_{2}^{e0}$ in the eigenvector of all emergent elastic waves differ from each other. It means that the lattice vibration of these modes has different magnitude in direction $x$ and direction $y$. On the other hand, when a tilted magnetic field is applied, all modes turn to emergent elastic waves at long wavelength limit, and the components of ${{{\mathbf{m}}^{c0}}}$ in the eigenvector of all modes are always nonzero, which means that all the vibrational modes can always be stimulated by an AC magnetic field. Our results show that the previously unexplained two novel magnon modes of SkX observed in $Cu_2OSeO_3$\cite{28} may indeed be induced by anisotropic effects. We theoretically prove that a tilted magnetic field will be very useful to tune the dynamical properties of the SkX. One can control the direction of anisotropy of dispersion relation by changing the direction of in-plane bias magnetic field; while greatly increasing the number of modes that can be excited by an AC magnetic field.

\section{Methods}
\
\\
$Free$ $energy$ $density$ $functional$ $of$ $B20$ $chiral$ $magnets$ $and$ $its$ $rescaling$\\
We use the following free energy density functional to study magnetic skyrmions in cubic helimagnets
\begin{equation}
\begin{aligned}
\phi \left( \mathbf{M} \right)=&\sum\limits_{i=1}^{3}{A{{\left( \frac{\partial \mathbf{M}}{\partial {{x}_{i}}} \right)}^{2}}+D}\mathbf{M}\cdot \left( \nabla \times \mathbf{M} \right)-\mathbf{B}\cdot \mathbf{M}\\&+\alpha (T-{{T}_{0}}){{\mathbf{M}}^{2}}+\beta {{\mathbf{M}}^{4}}+{{A}_{e}}\sum\limits_{i=1}^{3}{{{\left( \frac{\partial {{M}_{i}}}{\partial {{x}_{i}}} \right)}^{2}}},
\label{5}
\end{aligned}
\end{equation}
where $\mathbf{M}$ denotes the magnetization, $\mathbf{B}$ denotes the magnetic field, and $T$ denotes the temperature. The terms on the rhs. of eq. (\ref{5}) denote respectively the exchange energy density with a coefficient $A$, the Dzyaloshinskii-Moriya interaction (DMI) with a coefficient $D$, the Zeeman energy density, the second and fourth order Landau expansion terms, and the exchange anisotropy interaction with coefficient $A_e$. Eq. (\ref{5}) can be simplified by rescaling the spatial variables as $\mathbf{r}=\frac{\mathbf{x}}{{{L}_{D}}}$, $\mathbf{m}=\frac{\mathbf{M}}{{{M}_{0}}}$, $\mathbf{b}=\frac{\mathbf{B}}{{{B}_{0}}}$, $\tilde{T}=\frac{\alpha (T-{{T}_{0}})}{K}$, and ${{L}_{D}}=\frac{2A}{D}$, $K=\frac{{{D}^{2}}}{4A}$, ${{M}_{0}}=\sqrt{\frac{K}{\beta }}$, ${{B}_{0}}=2K{{M}_{0}}$, and $\tilde{A}_e=A_e/A$,which yield $\phi \left( \mathbf{M} \right)=\frac{{{K}^{2}}}{\beta }\tilde{\phi }\left( \mathbf{m} \right)$, while $\tilde{\phi }\left( \mathbf{m} \right)$ is given in eq. (\ref{1}). The benefit of eq. (\ref{1}) compared with eq. (\ref{5}) is that it provides a free energy density functional that is independent of material parameters, so that results obtained from analyzing eq. (\ref{1}) has general significance to Bloch-type SkX in any B20 helimagnets.
\
\\
\\
$Fourier$ $representation$ $of$ $SkX$ $and$ $free$ $energy$ $minimization$\\
To determine a metastable SkX state at given $\tilde{T}$ and $b$, one has to substitute the analytical expression of $\mathbf{m}$ for the SkX phase into eq. (\ref{1}), and minimize the free energy of the system with respect to all independent variables. In practice, we use the following Fourier representation of SkX instead of eq. (\ref{2})\cite{42}
\begin{equation}
    \begin{aligned}
        \mathbf m=& \mathbf m^c+\sum^n_{i=1}\sum^{n_i}_{j=1}\mathbf m_{\mathbf q_{ij}} e^{{\rm i}[\mathbf I-\mathbf F^e(\mathbf r)]^T\mathbf q_{ij}\cdot\mathbf r},
    \end{aligned}
    \label{9a}
\end{equation}
where $\mathbf m^c$ denotes a constant vector, $F_{ij}^e(\mathbf r)=\varepsilon^e_{ij}+\omega^e_{ij}=u^e_{i,j}$, and ${{\mathbf{q}}_{ij}}$ denote the undeformed wave vectors organized according to the following rules: $\left| {{\mathbf{q}}_{1j}} \right|<\left| {{\mathbf{q}}_{2j}} \right|<\left| {{\mathbf{q}}_{3j}} \right|<......$, and $\left| {{\mathbf{q}}_{i1}} \right|=\left| {{\mathbf{q}}_{i2}} \right|=......=\left| {{\mathbf{q}}_{i{{n}_{i}}}} \right|$. When truncated at a specific value of $n$, the $n^{th}$ order Fourier representation given in eq. (\ref{9a}) saves all the significant Fourier terms up to the $n^{th}$ order, which is hard to achieve if one uses eq. (\ref{2}). 

It is convenient to expand $\mathbf m_{\mathbf q_{ij}}$ as $\mathbf m_{\mathbf q_{ij}}=c_{ij1}\mathbf P_{ij1}+c_{ij2}\mathbf P_{ij2}+c_{ij3}\mathbf P_{ij3}$, where $c_{ij1}=c^{re}_{ij1}+{\rm i}c^{im}_{ij1}$, $c_{ij2}=c^{re}_{ij2}+{\rm i}c^{im}_{ij2}$, $c_{ij3}=c^{re}_{ij3}+{\rm i}c^{im}_{ij3}$ are complex variables to be determined, and $\mathbf P_{ij1}=\frac{1}{\sqrt{2}s_iq}[-{\rm i}q_{ijy}, {\rm i}q_{ijx}, s_iq]^T$, $\mathbf P_{ij2}=\frac{1}{s_iq}[q_{ijx}, q_{ijy}, 0]^T$, $\mathbf P_{ij3}=\frac{1}{\sqrt{2}s_iq}[{\rm i}q_{ijy}, -{\rm i}q_{ijx}, s_iq]^T$ with $\mathbf q_{ij}=[q_{ijx}, q_{ijy}]^T$, $|\mathbf q_{ij}|=s_iq$. For 2-D hexagonal SkX, we can assume without loss of generality that $\mathbf q_{11}=[0,1]^T$, $\mathbf q_{12}=[-\frac{\sqrt{3}}{2},-\frac{1}{2}]^T$, and $q=1$. Comparing eq. (\ref{9a}) and eq. (\ref{2}), we have ${\mathbf{q}}^e_{ij}(\varepsilon _{ij}^{e},\ {{\omega }^{e}})=[\mathbf I-(\mathbf F^e(\mathbf r))^T]{\mathbf{q}}_{ij}$, which gives for the basic reciprocal vectors $\mathbf q_{11}=[-\varepsilon _{12}^{e}+{\omega }^{e},1-\varepsilon _{22}^{e}]^T$, $\mathbf q_{12}=[-\frac{\sqrt{3}}{2}(1-\varepsilon _{11}^{e})+\frac{1}{2}(\varepsilon _{12}^{e}-{\omega }^{e}),\frac{\sqrt{3}}{2}(\varepsilon _{12}^{e}+{\omega }^{e})-\frac{1}{2}(1-\varepsilon _{22}^{e})]^T$.

In this case, all the independent variables describing the rescaled magnetization of the SkX phase can be gathered in two vectors, which are given by
\begin{equation}
    \begin{aligned}
    {{\pmb{\varepsilon}}^{ea}}={{\left[ \varepsilon _{11}^{e},\ \varepsilon _{22}^{e},\ \varepsilon _{12}^{e},\ \omega^{e} \right]}^{T}},
    \label{9b}
    \end{aligned}
\end{equation}
and
\begin{equation}
\begin{aligned}
{{\mathbf{m}}^{q}}=[& {{m}^c_{1}},\ {{m}^c_{2}},\ {{m}^c_{3}},c_{111}^{re},c_{112}^{re},c_{113}^{re},c_{121}^{re},c_{122}^{re},\\
& c_{123}^{re},c_{131}^{re},c_{132}^{re},c_{133}^{re},c_{111}^{im},c_{112}^{im},c_{113}^{im}\cdots]^{T},
\label{10}
\end{aligned}
\end{equation}
where the length of $\mathbf{m}^{q}$ depends on the order of Fourier representation used. For $1^{st}$, $2^{nd}$, and $3^{rd}$ order Fourier representation, the length of $\mathbf{m}^{q}$ is 21, 39 and 57, respectively. At given $\tilde{T}$ and $b$, minimization of the free energy based on eq. (\ref{9a}) determines the equilibrium value of the two vectors ${{\pmb{\varepsilon}}^{ea}}$ and $\mathbf{m}^{q}$ for a metastable SkX phase, which are denoted by $({{\pmb{\varepsilon}}^{ea}})_{st}$ and $(\mathbf{m}^{q})_{st}$. To see if the metastable SkX corresponds to the thermodynamically stable state, we have to compare the minimized free energy for the SkX phase with that for the other possible states such as the ferromagnetic phase, the G-conical phase, and the IPSQ phase. Expression of magnetization for the ferromagnetic phase and the IPSQ phase are included in eq. (\ref{9a}), while expression of the rescaled magnetization for the G-conical phase is composed of a constant vector plus a helix of magnetization, where both the direction of the constant vector and the wave vector of the helix are free to rotate in space.

When anisotropic effects are considered, the solution of $({{\pmb{\varepsilon}}^{ea}})_{st}$ and $(\mathbf{m}^{q})_{st}$ generally possesses lower symmetry. For example, for hexagonal SkX free from any in-plane anisotropy, we always have $({{{\varepsilon}}^{e}_{11}})_{st}=({{{\varepsilon}}^{e}_{22}})_{st}$ and $({{{\varepsilon}}^{e}_{12}})_{st}=(\omega^e)_{st}=0$. When anisotropic effects are presented, this condition is broken, and the four components of $({{\pmb{\varepsilon}}^{ea}})_{st}$ are all nonzero and different from each other. The change of values of $({{\pmb{\varepsilon}}^{ea}})_{st}$ and $(\mathbf{m}^{q})_{st}$ induced by anisotropy leads to drastic variation of the emergent phonons of SkX. 
\
\\
\\
$Euler-Lagrange$ $equation$ $for$ $the$ $emergent$ $displacements$ $and$ $the$ $Fourier$ $magnitudes$ $of$ $SkX$\\
Consider small amplitude vibration around the metastable or stable SkX phase, eq. (\ref{9a}) transforms to
\begin{equation}
    \begin{aligned}
        \mathbf m=& (\mathbf m^c)_{st}+(\mathbf m^c)_{v}+\sum^\infty_{i=1}\sum^{n_i}_{j=1}[(\mathbf m_{\mathbf q_{ij}})_{st}+(\mathbf m_{\mathbf q_{ij}})_{v}]\\
        &\times e^{{\rm i}[\mathbf I-(\mathbf F^e(\mathbf r))_{st}]^T\mathbf q_{ij}\cdot[\mathbf r-(\mathbf u^e)_{v}]},
    \end{aligned}
    \label{9}
\end{equation}
where $(F_{ij}^e(\mathbf r))_{st}=(\varepsilon^e_{ij})_{st}+(\omega^e_{ij})_{st}$.

The Euler-Lagrange equations of ${{\left( {{\mathbf{u}}^{e}} \right)}_{v}}$ and ${{\left( {{\mathbf{m}}^{q}} \right)}_{v}}$ read\cite{29, 44}
\begin{equation}
\begin{aligned}
-\frac{d}{dt}\left[ \frac{\partial \mathscr{L}}{\partial {{\left( {{{\mathbf{\dot{u}}}}^{e}} \right)}_{v}}} \right]+\frac{\partial \mathscr{L}}{\partial {{\left( {{\mathbf{u}}^{e}} \right)}_{v}}}-\sum\limits_{i}{\frac{d}{d{{r}_{i}}}\left[ \frac{\partial \mathscr{L}}{\partial {{\left( \mathbf{u}_{,i}^{e} \right)}_{v}}} \right]}=0,
\label{11}
\end{aligned}
\end{equation}
\begin{equation}
\begin{aligned}
-\frac{d}{dt}\left[ \frac{\partial \mathscr{L}}{\partial {{\left( {{{\mathbf{\dot{m}}}}^{q}} \right)}_{v}}} \right]+\frac{\partial \mathscr{L}}{\partial {{\left( {{\mathbf{m}}^{q}} \right)}_{v}}}-\sum\limits_{i}{\frac{d}{d{{r}_{i}}}\left[ \frac{\partial \mathscr{L}}{\partial {{\left( \mathbf{m}_{,i}^{q} \right)}_{v}}} \right]}=0,
\label{12}
\end{aligned}
\end{equation}
where $\mathscr{L}={{E}_{K}}-\Phi $ is the Lagrangian of the system, ${{E}_{K}}$ denotes the kinetic energy, $\Phi$ denotes the free energy, and $\mathbf{u}_{,i}^{e}=\frac{\partial {{\mathbf{u}}^{e}}}{\partial {{r}_{i}}}$. To actually use eqs. (\ref{11}, \ref{12}) for small vibration of ${{\left( {{\mathbf{u}}^{e}} \right)}_{v}}$ and ${{\left( {{\mathbf{m}}^{q}} \right)}_{v}}$, we first expand the averaged rescaled free energy density  $\bar{\phi }=\frac{1}{V}\int{\tilde{\phi }\left( \mathbf{m} \right)}dV$ in terms of ${{\left( {{\mathbf{u}}^{e}} \right)}_{v}}$ and ${{\left( {{\mathbf{m}}^{q}} \right)}_{v}}$ and their derivatives and retain the lowest order terms. Substituting ${{\left( {{\mathbf{u}}^{e}} \right)}_{v}}={{\mathbf{u}}^{e0}}{{e}^{\text{i}(\mathbf{\tilde{k}}\cdot \mathbf{r}-\omega t)}}$, ${{\left( {{\mathbf{m}}^{q}} \right)}_{v}}={{\mathbf{m}}^{q0}}{{e}^{\text{i}(\mathbf{\tilde{k}}\cdot \mathbf{r}-\omega t)}}$ into eqs. (\ref{11}, \ref{12}), we have
\begin{equation}
    \begin{aligned}
    \left( \mathbf{R}\omega -\mathbf{K} \right)\left[ \begin{matrix}
       {{\mathbf{u}}^{e0}}  \\
       {{\mathbf{m}}^{q0}}  \\
    \end{matrix} \right]=\mathbf{0},
    \label{4}
    \end{aligned}
    \end{equation}
where
\begin{equation}
\begin{aligned}
\mathbf{R}=\left[ \begin{matrix}
   {{\mathbf{R}}^{e}} & {{\mathbf{R}}^{eq}}  \\
   {{\left( {{\mathbf{R}}^{eq*}} \right)}^{T}} & {{\mathbf{R}}^{q}}  \\
\end{matrix} \right],
\label{13}
\end{aligned}
\end{equation}
\begin{equation}
\begin{aligned}
\mathbf{K }=\left[ \begin{matrix}
   {{\mathbf{K }}^{e}} & {{\mathbf{K }}^{eq}}  \\
   {{\left( {{\mathbf{K }}^{eq*}} \right)}^{T}} & {{\mathbf{K }}^{q}}  \\
\end{matrix} \right].
\label{14}
\end{aligned}
\end{equation}
Here ${{\mathbf{R}}^{eq*}}$ and ${{\mathbf{K }}^{eq*}}$ denote complex conjugate of ${{\mathbf{R}}^{eq}}$ and ${{\mathbf{K }}^{eq}}$. In eqs. (\ref{13}, \ref{14}), $R_{ij}^{e}=-\text{i}\frac{1}{V}{{\left[ \frac{\partial }{\partial \dot{u}_{j}^{e}}\left( \frac{\delta {{E}_{K}}}{\delta u_{i}^{e}} \right) \right]}_{st}},$ $R_{ij}^{q}= -\text{i}\frac{1}{V}{{\left[ \frac{\partial }{\partial \dot{m}_{j}^{q}}\left( \frac{\delta {{E}_{K}}}{\delta m_{i}^{q}} \right) \right]}_{st}},$ $R_{ij}^{eq}= -\text{i}\frac{1}{V}{{\left[ \frac{\partial }{\partial \dot{m}_{j}^{q}}\left( \frac{\delta {{E}_{K}}}{\delta u_{i}^{e}} \right) \right]}_{st}}$, $K _{ij}^{e}=\sum\limits_{p,s}{{{{\tilde{k}}}_{p}}{{{\tilde{k}}}_{s}}{{\left[ \frac{\partial }{\partial u_{j,ps}^{e}}\left( \frac{d}{d{{r}_{p}}}\left( \frac{\partial \bar{\phi }}{\partial u_{i,p}^{e}} \right) \right) \right]}_{st}}},$ $K _{ij}^{eq}={{\left[ \sum\limits_{p,s}{{{{\tilde{k}}}_{p}}{{{\tilde{k}}}_{s}}}\frac{\partial }{\partial m_{j,ps}^{q}}\left( \frac{d}{d{{r}_{p}}}\frac{\partial \phi }{\partial u_{i,p}^{e}} \right)-\sum\limits_{p}{\text{i}}{{{\tilde{k}}}_{p}}\frac{\partial }{\partial m_{j,p}^{q}}\left( \frac{d}{d{{r}_{p}}}\frac{\partial \phi }{\partial u_{i,p}^{e}} \right) \right]}_{st}},$ $ K _{ij}^{q}={{\left[ \frac{\partial }{\partial m_{j}^{q}}\left( \frac{\partial \bar{\phi }}{\partial m_{i}^{q}} \right)+\sum\limits_{p}{\text{i}}{{{\tilde{k}}}_{p}}\frac{\partial }{\partial m_{j,p}^{q}}\left( \frac{\partial \phi }{\partial m_{i}^{q}} \right)-\sum\limits_{p}{\text{i}}{{{\tilde{k}}}_{p}}\frac{\partial }{\partial m_{j,p}^{q}}\right.}}$ ${{\left.\left( \frac{d}{d{{r}_{p}}}\left( \frac{\partial \phi }{\partial m_{i,p}^{q}} \right) \right)+\sum\limits_{p,s}{{{{\tilde{k}}}_{p}}{{{\tilde{k}}}_{s}}}\frac{\partial }{\partial m_{j,ps}^{q}}\left( \frac{d}{d{{r}_{p}}}\left( \frac{\partial \phi }{\partial m_{i,p}^{q}} \right) \right) \right]}_{st}}.$ Here a subscript $"st"$ means that the term is calculated at the equilibrium state ${{\mathbf{u}}^{e}}={{\left( {{\mathbf{u}}^{e}} \right)}_{st}}$ and ${{\mathbf{m}}^{q}}={{\left( {{\mathbf{m}}^{q}} \right)}_{st}}$. One should notice that the stiffness matrix $\mathbf{K}$ is completely determined by the emergent elasticity of the SkX under magnetic field\cite{42,44}. The dispersion relation for different modes ${{\omega }_{i}}={{\omega }_{i}}(\mathbf{\tilde{k}})$ can be obtained by solving eq. (\ref{4}). By defining the rescaled frequency ${{\tilde{\omega }}_{i}}=\frac{1}{\eta }{{\omega }_{i}}$, where $\eta =\left| \frac{\gamma {{m}^{3}}{{D}^{4}}}{16{{M}}{{A}^{2}}\beta } \right|$, one obtains a rescaled dispersion relation ${{\tilde{\omega }}_{i}}={{\tilde{\omega }}_{i}}(\mathbf{\tilde{k}})$ that is material-independent.

\begin{acknowledgments}
The work was supported by the NSFC (National Natural Science Foundation of China) through the funds 11772360, 11472313, 11572355 and Pearl River Nova Program of Guangzhou (Grant No. 201806010134).
\end{acknowledgments}

\bibliographystyle{apsrev4-1}

\begin{thebibliography}{99}
    \bibitem{1} 	M$\rm\ddot{u}$hlbauer, S. et al. Skyrmion Lattice in a Chiral Magnet. Science \textbf{323} 915–919 (2009).
    \bibitem{2} 	Yu, X. Z. et al. Real-space observation of a two-dimensional skyrmion crystal. Nature \textbf{465} 901–904 (2010).
    \bibitem{3} 	Karube, K. et al. Robust metastable skyrmions and their triangular-square lattice structural transition in a high-temperature chiral magnet. Nature Materials \textbf{15} 1237–1242 (2016).
    \bibitem{4}     Bogdanov, A. \& Hubert, A. Thermodynamically stable magnetic vortex states in magnetic crystals. Journal of Magnetism and Magnetic Materials \textbf{138} 255-269 (1994).
    \bibitem{5} 	Bogdanov, A. N., Pfleiderer, C. \& R$\rm\ddot{o}$ßler, U. K. Spontaneous skyrmion ground states in magnetic metals. Nature \textbf{442} 797 (2006).
    \bibitem{6} 	Garst, M., Waizner, J. \& Grundler, D. Collective spin excitations of helices and magnetic skyrmions: review and perspectives of magnonics in non-centrosymmetric magnets. Journal of Physics D-Applied Physics \textbf{50} 293002 (2017).
    \bibitem{7} 	Heinze, S. et al. Spontaneous atomic-scale magnetic skyrmion lattice in two dimensions. Nature Physics \textbf{7} 713–718 (2011).
    \bibitem{8} 	Soumyanarayanan, A. et al. Tunable room-temperature magnetic skyrmions in Ir/Fe/Co/Pt multilayers. Nature Materials \textbf{16} 898-+ (2017).
    \bibitem{9} 	Oike, H. et al. Interplay between topological and thermodynamic stability in a metastable magnetic skyrmion lattice. Nature Physics \textbf{12} 62–66 (2016).
    \bibitem{10} 	Hu, Y. Wave nature and metastability of emergent crystals in chiral magnets. Communications Physics \textbf{1} 82 (2018).
    \bibitem{11} 	Shibata, K. et al. Large anisotropic deformation of skyrmions in strained crystal. Nature Nanotechnology \textbf{10} 589–592 (2015).
    \bibitem{12} 	Morikawa, D. et al. Deformation of Topologically-Protected Supercooled Skyrmions in a Thin Plate of Chiral Magnet Co8Zn8Mn4. Nano Letters \textbf{17} 1637–1641 (2017).
    \bibitem{45} 	Hu, Y., Lan, X. \& Wang, B. Nonlinear emergent elasticity and structural transitions of skyrmion crystal under uniaxial distortion. arXiv:1812.06311 [cond-mat.mes-hall] (2018).
    \bibitem{13} 	Nakajima, T. et al. Skyrmion lattice structural transition in MnSi. Science advances \textbf{3} e1602562–e1602562 (2017).
    \bibitem{14} 	Wang, C. et al. Enhanced Stability of the Magnetic Skyrmion Lattice Phase under a Tilted Magnetic Field in a Two-Dimensional Chiral Magnet. Nano Letters \textbf{17} 2921–2927 (2017).
    \bibitem{15} 	White, J. S. et al. Electric-Field-Induced Skyrmion Distortion and Giant Lattice Rotation in the Magnetoelectric Insulator Cu2OSeO3. Physical Review Letters \textbf{113} 107203 (2014).
    \bibitem{16} 	Onose, Y., Okamura, Y., Seki, S., Ishiwata, S. \& Tokura, Y. Observation of Magnetic Excitations of Skyrmion Crystal in a Helimagnetic Insulator Cu2OSeO3. Physical Review Letters \textbf{109} 037603 (2012).
    \bibitem{17} 	Buettner, F. et al. Dynamics and inertia of skyrmionic spin structures. Nature Physics \textbf{11} 225–228 (2015).
    \bibitem{18} 	Makhfudz, I., Krueger, B. \& Tchernyshyov, O. Inertia and Chiral Edge Modes of a Skyrmion Magnetic Bubble. Physical Review Letters \textbf{109} 217201 (2012).
    \bibitem{19} 	Schwarze, T. et al. Universal helimagnon and skyrmion excitations in metallic, semiconducting and insulating chiral magnets. Nature Materials \textbf{14} 478–483 (2015).
    \bibitem{20} 	Takagi, R. et al. Spin-wave spectroscopy of the Dzyaloshinskii-Moriya interaction in room-temperature chiral magnets hosting skyrmions. Physical Review B \textbf{95} 220406 (2017).
    \bibitem{21} 	Mochizuki, M. Spin-Wave Modes and Their Intense Excitation Effects in Skyrmion Crystals. Physical Review Letters \textbf{108} 017601 (2012).
    \bibitem{22} 	Petrova, O. \& Tchernyshyov, O. Spin waves in a skyrmion crystal. Phys. Rev. B \textbf{84} 214433 (2011).
    \bibitem{23} 	Langner, M. C. et al. Nonlinear Ultrafast Spin Scattering in the Skyrmion Phase of Cu2OSeO3. Physical Review Letters \textbf{119} 107204 (2017).
    \bibitem{24} 	Gomonay, O., Baltz, V., Brataas, A. \& Tserkovnyak, Y. Antiferromagnetic spin textures and dynamics. Nature Physics \textbf{14} 213 (2018).
    \bibitem{25} 	Bak, P. \& Jensen, M. H. Theory of helical magnetic structures and phase transitions in MnSi and FeGe. J. Phys. C: Solid State Phys. \textbf{13} L881 (1980).
    \bibitem{26} 	Luo, Y. et al. Anisotropic magnetocrystalline coupling of the skyrmion lattice in MnSi. Phys. Rev. B \textbf{97} 104423 (2018).
    \bibitem{27} 	Wan, X., Hu, Y. \& Wang, B. Exchange-anisotropy-induced intrinsic distortion, structural transition, and rotational transition in skyrmion crystals. Phys. Rev. B \textbf{98} 174427 (2018).
    \bibitem{28} 	Tucker, G. S. et al. Spin excitations in the skyrmion host Cu2OSeO3. Physical Review B \textbf{93} 054401 (2016).
    \bibitem{29} 	Hu, Y. Emergent elastic waves in skyrmion crystals with finite frequencies at long wavelength limit. ArXiv (2019).
    \bibitem{30} 	Lin, S.-Z. \& Saxena, A. Noncircular skyrmion and its anisotropic response in thin films of chiral magnets under a tilted magnetic field. Physical Review B \textbf{92} 180401 (2015).
    \bibitem{31} 	Leonov, A. O. \& Kézsmárki, I. Skyrmion robustness in non-centrosymmetric magnets with axial symmetry: The role of anisotropy and tilted magnetic fields. Physical Review B \textbf{96} 214413 (2017).
    \bibitem{32} 	Kezsmarki, I. et al. Neel-type skyrmion lattice with confined orientation in the polar magnetic semiconductor GaV4S8. Nature Materials \textbf{14} 1116-+ (2015).
    \bibitem{33} 	Wan, X., Hu, Y. \& Wang, B. Controlling stability and emergent rotation of the skyrmion crystal in thin films of helimagnets via tilted magnetic field. Phys. Rev. B (2019).
    \bibitem{34} 	Zang, J., Mostovoy, M., Han, J. H. \& Nagaosa, N. Dynamics of Skyrmion Crystals in Metallic Thin Films. Phys. Rev. Lett. \textbf{107} 136804 (2011).
    \bibitem{35} 	Schulz, T. et al. Emergent electrodynamics of skyrmions in a chiral magnet. Nature Physics \textbf{8} nphys2231 (2012).
    \bibitem{36} 	Jonietz, F. et al. Spin Transfer Torques in MnSi at Ultralow Current Densities. Science \textbf{330} 1648–1651 (2010).
    \bibitem{37} 	Woo, S. et al. Spin-orbit torque-driven skyrmion dynamics revealed by time-resolved X-ray microscopy. Nature Communications \textbf{8} 15573 (2017).
    \bibitem{38} 	Neubauer, A. et al. Topological Hall Effect in the A Phase of MnSi. Physical Review Letters \textbf{102} 186602 (2009).
    \bibitem{39} 	Franz, C. et al. Real-Space and Reciprocal-Space Berry Phases in the Hall Effect of Mn1-xFexSi. Physical Review Letters \textbf{112} 186601 (2014).
    \bibitem{40} 	Litzius, K. et al. Skyrmion Hall effect revealed by direct time-resolved X-ray microscopy. Nature Physics \textbf{13} 170–175 (2017).
    \bibitem{41} 	Jiang, W. et al. Direct observation of the skyrmion Hall effect. Nature Physics \textbf{13} 162–169 (2017).
    \bibitem{42} 	Hu, Y. \& Wan, X. Thermodynamics and elasticity of emergent crystals. arXiv: 1905.02165 (2019).
    \bibitem{43} 	Yu, X. Z. et al. Near room-temperature formation of a skyrmion crystal in thin-films of the helimagnet FeGe. Nature Materials \textbf{10} 106–109 (2011).
    \bibitem{44} 	Hu, Y. Lagrangian formulation for emergent elastic waves in magnetic emergent crystals. ArXiv (2019).    
\end{thebibliography}

\newpage
\begin{figure*}
\centering
\includegraphics[scale=0.3]{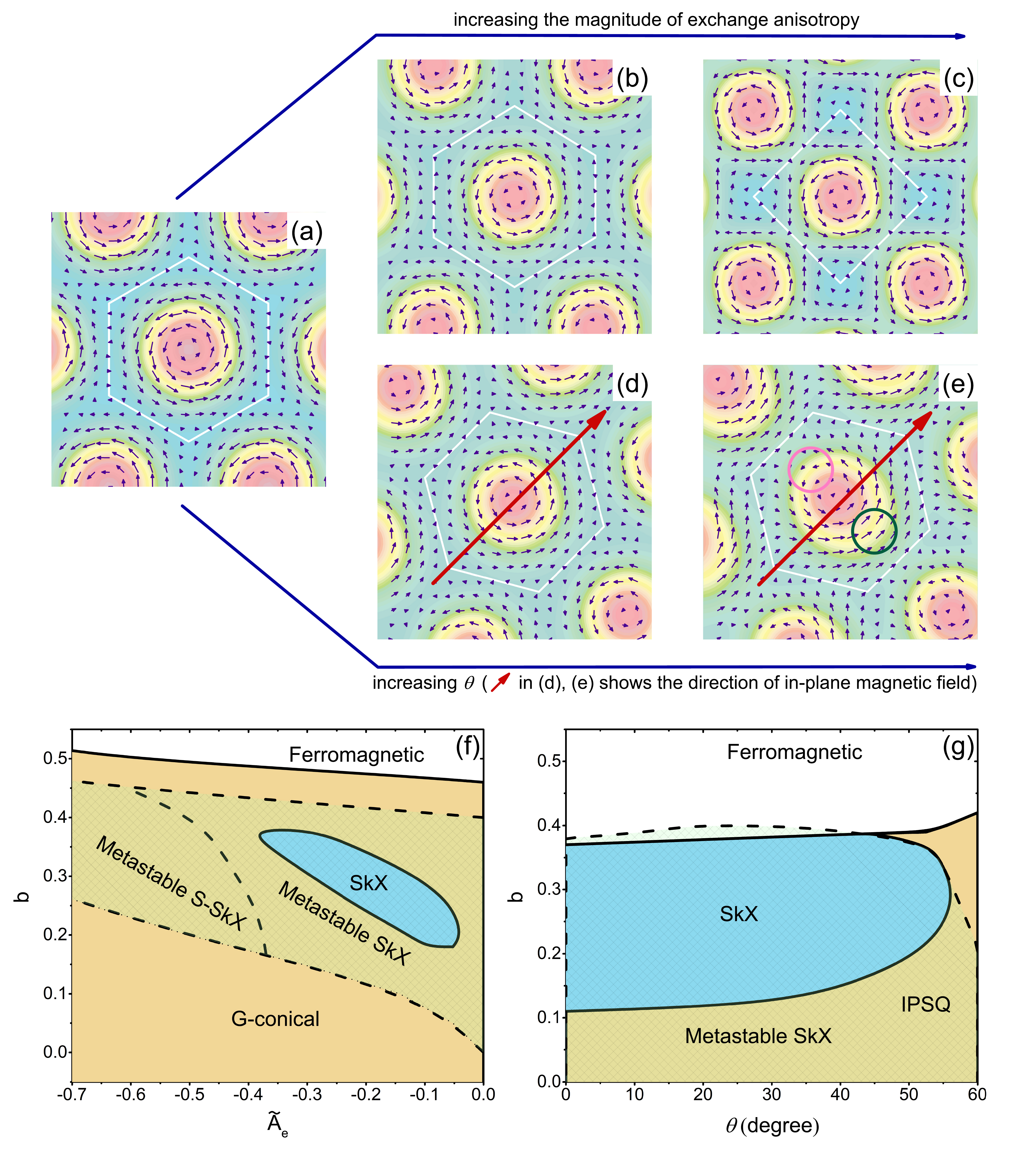}
\caption{\label{fig1}Variation of the equilibrium properties of the SkX phase with different anisotropic effects. (a-c)Variation of the field pattern of SkX with exchange anisotropy, where (a) is plotted at $\tilde{A}_e=0, \theta=0^o$, (b) is plotted at $\tilde{A}_e=-0.15, \theta=0^o$, and (c) is plotted at $\tilde{A}_e=-0.6, \theta=0^o$. (a, d, e)Variation of the field pattern of SkX with tilted angle of bias magnetic field, where (d) is plotted at $\tilde{A}_e=0, \theta=15^o$, and (e) is plotted at $\tilde{A}_e=0, \theta=45^o$. (f) $\tilde{A}_e-b$ phase diagram and (g) $\theta-b$ phase diagram of helimagnets. In the two phase diagrams, the stable SkX region is marked blue while the metastable SkX region corresponds to a shadowed light green area.}
\end{figure*}

\begin{figure*}
\includegraphics[scale=0.26]{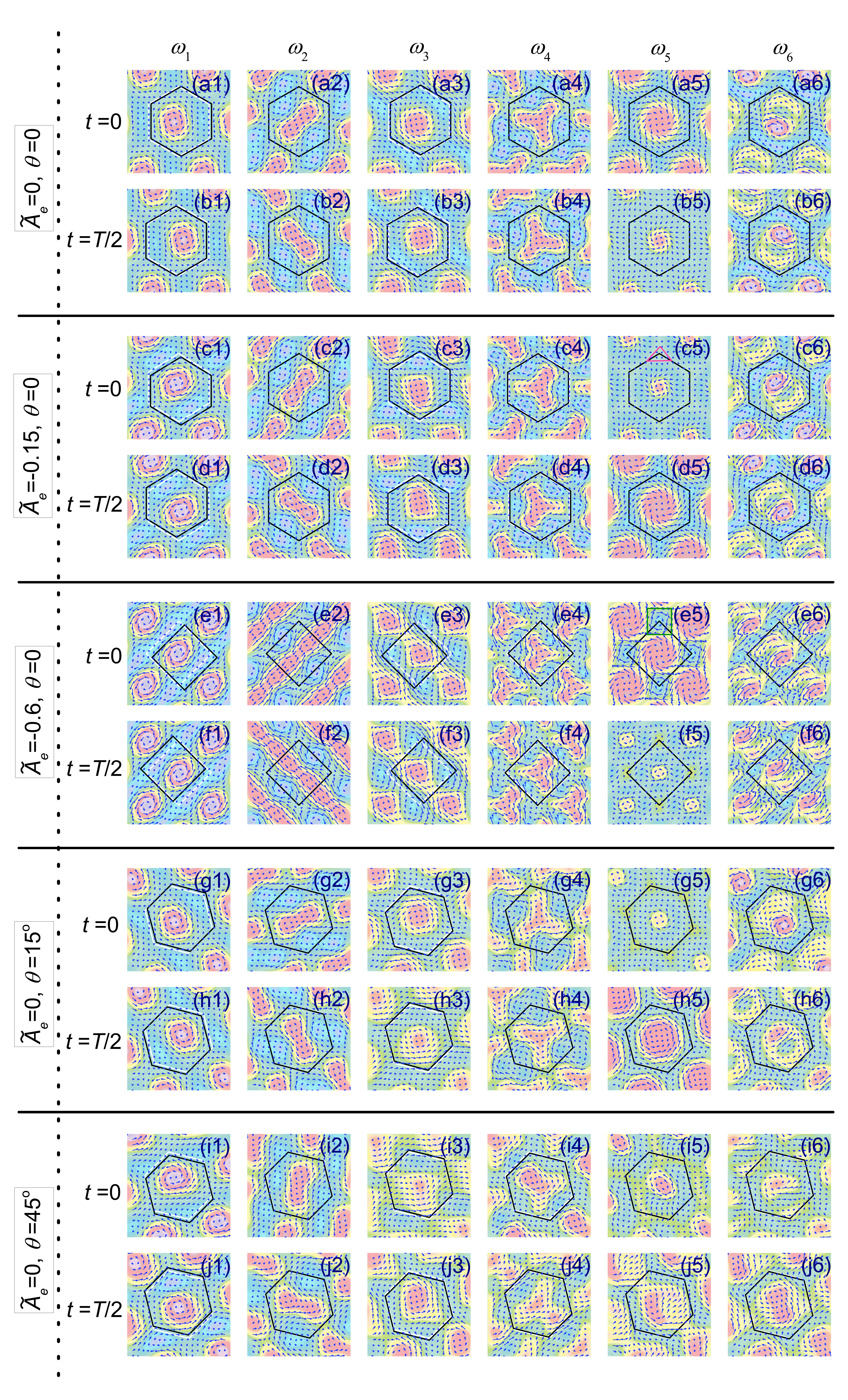}
\caption{\label{fig2}Field configurations of SkX undergoing the first 6 modes of emergent phonons calculated at $\tilde{T}=0.5,\ b=0.3$ near the $\Gamma $ point (${{\tilde{k}}_{1}}={{10}^{-\text{5}}},\ {{\tilde{k}}_{2}}=0$), where five different conditions of anisotropy are considered: (a$s$, b$s$) $\tilde{A}_e=0, \theta=0$, (c$s$, d$s$)  $\tilde{A}_e=-0.15, \theta=0$, (e$s$, f$s$)  $\tilde{A}_e=-0.6, \theta=0$, 
(g$s$, h$s$)  $\tilde{A}_e=-0, \theta=15^o$, 
(i$s$, j$s$) $\tilde{A}_e=-0, \theta=45^o$, ($s=1, 2, \cdots,6$). 
Different plots at two time points $t$ during a period $T$ are shown. In all figures, the vectors illustrate the distribution of the in-plane magnetization components with length proportional to their magnitude, while the colored density plot illustrates the distribution of the out-of-plane magnetization component. The black solid line plots the displaced Wigner-Seitz cell of SkX due to wave motion, while the white solid line plots the static position of the Wigner-Seitz cell.} 
\end{figure*}

\begin{figure}
\includegraphics[scale=0.3]{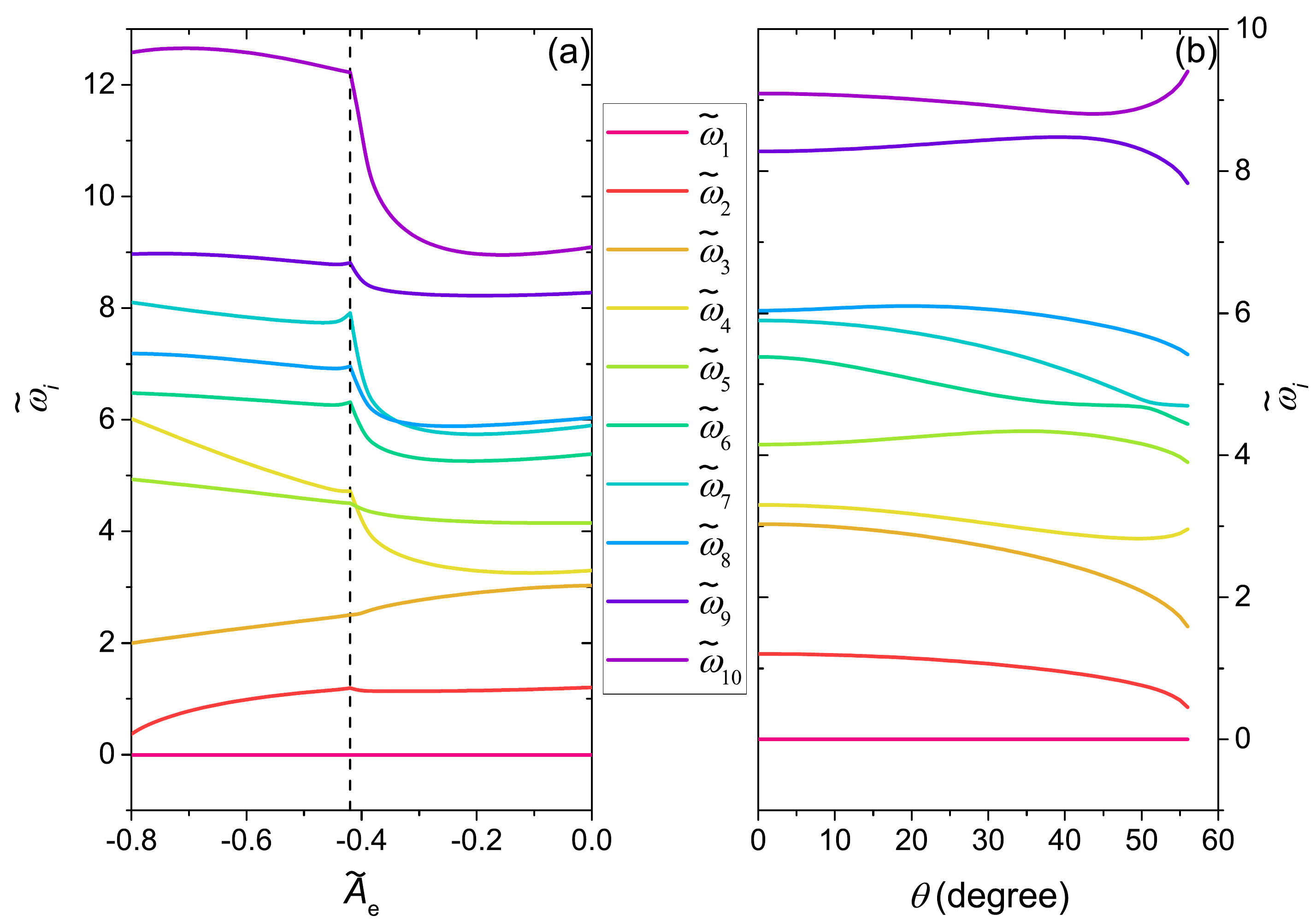}
\caption{\label{fig3} Variation of the frequencies of the first 10 modes of emergent phonons of SkX with (a) $\tilde{A}_e$, (b) $\theta$ calculated at $\tilde{T}=0.5,\ b=0.3$ near the $\Gamma $ point (${{\tilde{k}}_{1}}={{10}^{-\text{5}}},\ {{\tilde{k}}_{2}}=0$).} 
\end{figure}

\begin{figure*}
   \includegraphics[scale=0.36]{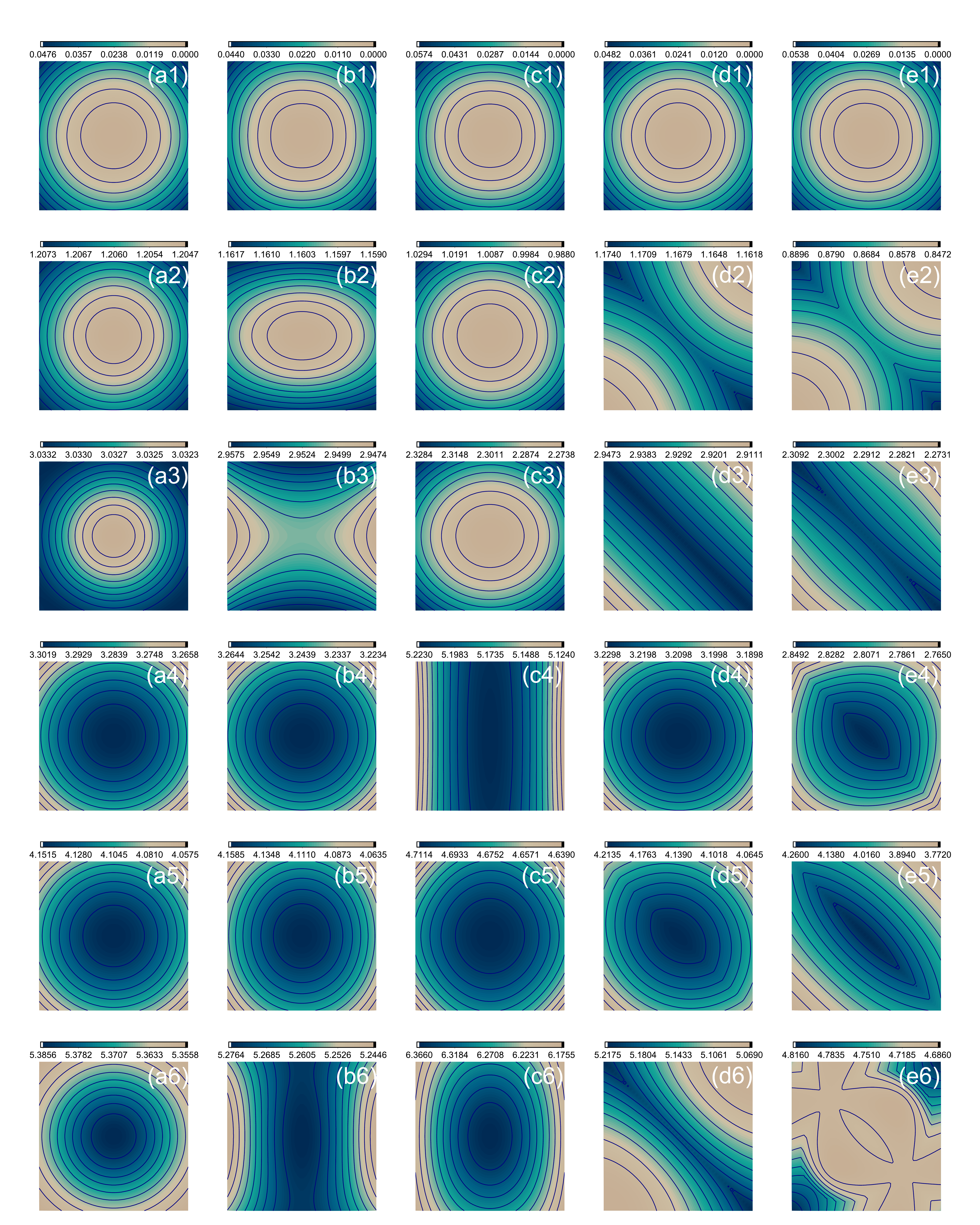}
   \caption{\label{fig4}Contours of equi-frequency with constant spacing in the reciprocal space for the first six vibrational modes calculated at five different conditions of aniostroy: (a$s$) $\tilde{A}_e=0, \theta=0$, (b$s$)  $\tilde{A}_e=-0.15, \theta=0$, (c$s$)  $\tilde{A}_e=-0.6, \theta=0$, 
   (d$s$)  $\tilde{A}_e=-0, \theta=15^o$, and
   (e$s$) $\tilde{A}_e=-0, \theta=45^o$, ($s=1, 2, \cdots,6$), where the $\Gamma $ point is in the center of each plot, and ${{\tilde{k}}_{1}}\in [-0.5,\ 0.5],\ {{\tilde{k}}_{2}}\in [-0.5,\ 0.5]$.} 
\end{figure*}

\begin{table*}
    \caption{Specific component of the eigenvectors of the first sixteen modes calculated at $t=0.5$, $b=0.3$ with five different conditions of anisotropy.}
\begin{tabular}{c|l|cccccccccccccccccccc} 
\toprule 
&Modes &$\omega_{1}$ &$\omega_{2}$ &$\omega_{3}$ &$\omega_{4}$ &$\omega_{5}$ &$\omega_{6}$ &$\omega_{7}$ &$\omega_{8}$ &$\omega_{9}$ &$\omega_{10}$ &$\omega_{11}$ &$\omega_{12}$ &$\omega_{13}$ &$\omega_{14}$ &$\omega_{15}$ &$\omega_{16}$ 
\\
\hline
\midrule \multirowcell{4}{$\tilde{A}_e=0$\\ $\theta=0^{\rm o}$}&$u^{e0}_{1}$ &0.105 &0 &0.088 &0 &0 &0.011 &0 &0 &0 &0.014 &0 &0 &0.018 &0.001 &0 &0\\\cline{2-18} 
&$u^{e0}_{2}$ &0.178 &0 &0.088 &0 &0 &0.011 &0 &0 &0 &0.014 &0 &0 &0.018 &0.001 &0 &0\\\cline{2-18} 
&$m^{c0}_1$ ($m^{c0}_2$)  &0 &0 &0.532 &0 &0 &0.167 &0 &0 &0 &0.158 &0 &0 &0.214 &0.063 &0 &0\\\cline{2-18} 
&$m^{c0}_3$  &0 &0 &0 &0 &0.558 &0 &0 &0 &0 &0 &0 &0.138 &0 &0 &0.281 &0\\ 
\hline
\midrule \multirowcell{5}{$\tilde{A}_e=-0.15$\\ $\theta=0^{\rm o}$}
&$u^{e0}_{1}$ &0.114 &0 &0.092 &0.042 &0 &0.011 &0 &0 &0.007 &0.012 &0 &0 &0.019 &0.004 &0 &0\\\cline{2-18} 
&$u^{e0}_{2}$ &0.185 &0 &0.091 &0.038 &0 &0.014 &0 &0 &0.003 &0.016 &0 &0 &0.018 &0.005 &0 &0\\\cline{2-18} 
&$m^{c0}_1$ &0 &0 &0.499 &0.247 &0 &0.186 &0 &0 &0.035 &0.167 &0 &0 &0.179 &0.078 &0 &0\\\cline{2-18} 
&$m^{c0}_2$ &0 &0 &0.500 &0.239 &0 &0.161 &0 &0 &0.061 &0.155 &0 &0 &0.228 &0.033 &0 &0\\\cline{2-18} 
&$m^{c0}_3$ &0 &0.013 &0 &0 &0.569 &0 &0.014 &0.013 &0 &0 &0.053 &0.125 &0 &0 &0.259 &0.015\\ 
\hline
\midrule \multirowcell{5}{$\tilde{A}_e=-0.6$\\ $\theta=0^{\rm o}$}
&$u^{e0}_{1}$ &0.213 &0 &0.150 &0.042 &0 &0.007 &0 &0 &0.032 &0 &0.023 &0.018 &0 &0 &0.025 &0.017\\\cline{2-18} 
&$u^{e0}_{2}$ &0.196 &0 &0.154 &0.025 &0 &0.025 &0 &0 &0.031 &0 &0.030 &0.017 &0 &0 &0.007 &0.003\\\cline{2-18} 
&$m^{c0}_1$ &0 &0 &0.390 &0.356 &0 &0.241 &0 &0 &0.128 &0 &0.156 &0.059 &0 &0 &0.046 &0.116\\\cline{2-18} 
&$m^{c0}_2$ &0 &0 &0.377 &0.369 &0 &0.175 &0 &0 &0.321 &0 &0.152 &0.154 &0 &0 &0.030 &0.128\\\cline{2-18} 
&$m^{c0}_3$ &0 &0.011 &0 &0 &0.650 &0 &0.037 &0.037 &0 &0.292 &0 &0 &0.008 &0.261 &0 &0\\ 
\hline
\midrule \multirowcell{5}{$\tilde{A}_e=0$\\ $\theta=15^{\rm o}$}
&$u^{e0}_{1}$ &0.104 &0.007 &0.088 &0.006 &0.013 &0.012 &0.005 &0.005 &0.001 &0.013 &0.001 &0.006 &0.018 &0.005 &0.007 &0.002\\\cline{2-18} 
&$u^{e0}_{2}$ &0.177 &0.007 &0.088 &0.006 &0.013 &0.012 &0.005 &0.005 &0.001 &0.013 &0.001 &0.006 &0.018 &0.005 &0.007 &0.002\\\cline{2-18} 
&$m^{c0}_1$ &0 &0.023 &0.511 &0.042 &0.128 &0.148 &0.056 &0.085 &0.017 &0.136 &0.091 &0.076 &0.193 &0.050 &0.022 &0.105\\\cline{2-18} 
&$m^{c0}_2$ &0 &0.023 &0.511 &0.042 &0.128 &0.148 &0.056 &0.085 &0.017 &0.136 &0.091 &0.076 &0.193 &0.050 &0.022 &0.105\\\cline{2-18} 
&$m^{c0}_3$ &0 &0.001 &0.034 &0 &0.560 &0.089 &0.004 &0.035 &0 &0.045 &0.001 &0.111 &0.055 &0.123 &0.272 &0.037\\ 
\hline
\midrule \multirowcell{5}{$\tilde{A}_e=0$\\ $\theta=45^{\rm o}$}
&$u^{e0}_{1}$ &0.104 &0.018 &0.089 &0.053 &0.012 &0.025 &0.010 &0.018 &0.003 &0.015 &0.008 &0.018 &0.014 &0.024 &0.012 &0.013\\\cline{2-18} 
&$u^{e0}_{1}$ &0.166 &0.018 &0.089 &0.053 &0.012 &0.025 &0.010 &0.018 &0.003 &0.015 &0.008 &0.018 &0.014 &0.024 &0.012 &0.013\\\cline{2-18} 
&$m^{c0}_1$ &0 &0.074 &0.361 &0.289 &0.169 &0.191 &0.103 &0.177 &0.075 &0.087 &0.093 &0.180 &0.096 &0.039 &0.043 &0.163\\\cline{2-18} 
&$m^{c0}_2$ &0 &0.074 &0.361 &0.289 &0.169 &0.191 &0.103 &0.177 &0.075 &0.087 &0.093 &0.180 &0.096 &0.039 &0.043 &0.163\\\cline{2-18} 
&$m^{c0}_3$ &0 &0.023 &0.020 &0.033 &0.378 &0.486 &0.016 &0.263 &0.031 &0.113 &0.058 &0.041 &0.025 &0.302 &0.139 &0.241\\ 
\hline

\bottomrule 
\end{tabular} 
\end{table*}

\end{document}